\documentclass[12pt]{article}
\usepackage[english]{babel}
\usepackage{caption} 
\usepackage[dvipsnames,svgnames,x11names]{xcolor}
\usepackage[margin=1.5cm]{geometry}
\usepackage{graphicx} 
\usepackage{amsmath}
\usepackage{setspace}
\usepackage{adjustbox}

\title{Predicting the Acoustic Signatures of Saturn's Upper Atmosphere}\author{
Andrew Powell$^\bullet$, Andi Petculescu$^\bullet$,\\
Rishabh Chaudhary$^\dagger$, Robert White$^\dagger$,\\
Don Banfield$^\ddagger$, Ian Neeson$^\mathsection$\\[1em]
$^\bullet$ University of Louisiana at Lafayette (Lafayette, LA)\\
$^\dagger$ Tufts University (Medford, MA)\\
$^\ddagger$ NASA-Ames Research Center (Mountain View, CA)\\
$^\mathsection$ VN Instruments, Ltd (Brockville, ON, Canada)
}

\begin{document}
\maketitle
\begin{abstract}

{Predictions for the acoustic attenuation coefficient and phase speed as functions of frequency and altitude in Saturn's atmosphere are presented and discussed. The pressure range considered in the study is 1 mbar to 1 bar, in windless and cloudless conditions. The atmospheric composition is represented by the major constituents, namely hydrogen (with its two spin isomers, ortho-H$_2$ and para-H$_2$) and helium. The H$_2$ and He concentrations are assumed constant with respect to altitude; however, non-uniform ortho- and para-H$_2$ profiles are considered. The acoustic wavenumber is obtained by incorporating a viscous, thermal, and internal molecular relaxation effects in a linearized fluid dynamics model. The ambient inputs are vertical profiles of the specific heats, shear viscosity, and thermal conductivity coefficients of the three-component (oH$_2$, pH$_2$, He) mixture, extracted at each pressure-temperature pair. The authors acknowledge funding from NASA-Ames Center for Innovation Fund (CIF).}
\end{abstract}
\section{Introduction}

Planetary atmospheres are characterized by the interplay between chemical composition, meteorological dynamics, mass-energy transport mechanisms, thermal profiles, cloud formation etc. whose knowledge may lead to insights into their formation. 

As stated in NASA's Vision for Planetary Science in the Decade 2013-2022 \cite{vision2013planetary}, there is continued interest in the exploration of gas giants. 

To a certain extent, gas giants are bellwethers of the terrestrial planets' evolution, given that their different formation scenarios enable understanding of the inner planets' origins. In particular, the presence of Saturn and Jupiter led to the stability of the orbits of the terrestrial planets. This is one prominent result of the grand tack hypothesis\cite{Walsh2011}. Additionally, the gas giants provide a shield from impacts. Furthermore, thousands of exoplanets have been discovered. Currently, gas giants in the Solar System act as analogues to understand the many exoplanets being discovered at an increasing rate, notably since recent observations have detected many giant planets\cite{winn2023little}.  

The success of Cassini-Huygens has contributed to the production of many impactful studies\cite{saturn_from_cassini}. However, there remains many open questions. Due to this, the NASA New Frontiers Program has designated an eminent need for in-situ measurements of Saturn's atmosphere\cite{atkinson2017new}. It is towards this goal that the work contained here is framed. 
\par The accomplishments from the Juno mission to Jupiter has inspired a similar style of mission to the ringed planet. Whereas many in-situ measurements provided by the Galileo mission to Jupiter have been studied\cite{vonZahn_1998}\cite{Niemann_1998}, there has been no such attempt for Saturn. 

\subsection{Saturn's atmosphere}
The dominant component of the chemical species on Saturn is molecular hydrogen, followed by helium and methane. Together, these account for nearly 99.9\% of its atmosphere by volume\cite{saturn_from_cassini}. Therefore, examination of Saturn's atmosphere will require knowledge of these species. In particular, their chemical abundances will have implications that constrain formation and evolution models.

Saturn has a noticeable enrichment of heavy elements relative to the solar average\cite{CONRATH2000124}\cite{KOSKINEN2018161}. These observations yield pivotal constraints for formation scenarios. As processes that fractionate heavy elements are  less sensitive to remote observations, a new undertaking must be used to chisel out the space of possible scenarios.

\subsection{Ortho/para-hydrogen ratio}
The most abundant molecule in the Saturnian atmosphere is molecular hydrogen. It occurs in two spin isomers: ortho-hydrogen (oH$_2$) and para-hydrogen (pH$_2$), whose nuclear spins are parallel and anti-parallel, respectively. The prominent difference in the two nuclear isomers are in their respective rotational energy transitions. Transitions in ortho-hydrogen occur between odd-numbered energy levels ($J = 1,3,5,...$), while those in para-hydrogen are between even-numbered levels ($J = 0,2,4...$).


The difference in the transitions directly determines the rotational energy, and thus the collision rates of state to state energy exchange. There are many types of excitation, due to energy transfer, between these molecules. Predominantly, the collisions between the species are the energy transfers of interest. These include: oH2:oH2, pH2:pH2, and oH2:pH2. Molecular relaxation experiments conducted through molecular beam experiments provide the state to state rates of these collisions that allow the determination of the relaxation time\cite{montero2014rotational}.  

\subsection{Helium abundance}
One large unanswered question that has been elusive for many years is the abundance of helium present in Saturn. Many researchers have used occultation data from Voyager to place bounds on the helium content\cite{CONRATH2000124}. However, this is not a direct observation of the helium content, rather an estimation of the He-H2 collision induced absorption at $200-600 \mathrm{cm}^{-1}$ . Most results display a modest depletion in the observed helium abundance as compared to the solar and Jovian counterparts\cite{KOSKINEN2018161}.

During the formation of gas giant planets, hydrogen and helium are accreted from the proto-solar nebula. As a consequence, the mixing ratio for the species should fall within that of the primitive value. In the process of formation, different fractionations can occur to modify this distribution. Thus, any deviation would highly constrain potential planetary evolution models. Hence, measurements of the different concentrations of H2/He would help to examine which models are acceptable. Considering the formation of the giant planets occurs past the snow line, the volatile content could help to aggregate further growth and complex processes could sequester heavier elements deep within the interior. 

This mystery has been addressed with many different explanations, such as the immiscibility of helium to be partitioned to the core, as well as the potential for helium to be rained out of Saturn's atmosphere\cite{Wilson_2010}.

\section{Methodology}
\subsection{Planetary acoustics}
\par 
The relevant acoustic quantity that is sensitive to the atmospheric specification is the wavenumber. Its real and imaginary parts are related, respectively, to the phase speed (or speed of sound) and the attenuation coefficient. For example, in the absence of molecular relaxation, the sound speed is a direct thermodynamic footprint of the ambient temperature and mean molecular mass. In polyatomic molecules, internal relaxation processes become important (e.g. vibrational relaxation in large molecules and rotational relaxation in H$_2$), leaving their mark on both the sound speed and attenuation coefficient. Therefore, the ability to measure the two components of the wavenumber over specific frequency ranges enables the development of fast and rugged spacecraft-borne atmospheric sensors. This has been amply proven in recent years, exemplified by the Cassini-Huygens mission to Titan, as well as the Mars InSight and Mars 2020 missions that have already acquired an impressive amount of acoustic data. In the (near) future, the Dragonfly mission to Titan will also have an acoustic sensor package. 
\par

A nascent and growing subdiscipline within the field of acoustics is its application to planetary bodies\cite{petculescu2007atmospheric}. The field of planetary acoustics incorporates many different physical phenomena. The wildly diverse range of planetary environments serve as a unique setting for acoustic observations. From seismic coupling due to geological events, to meteorological phenomena such as thunder, a vast space is open to explore using acoustic techniques. 

\par Recently, an enormous success was the creation of the first anemometer for Mars\cite{banfield2016martian}. The use of time of flight differences provides high accuracy in determining the sound speed\cite{white2023martian}. A similar approach could be applied to future instruments that are sent to Saturn.

\subsection{Rotational relaxation}
\par From the vantage point of Kinetic Theory, molecular collisions yield energy transfer\cite{hirschfelder1964molecular}. Characteristic energy transfer is uniquely determined from the species involved in the collisions. This energy transfer allows for relaxation of the internal and external degrees of freedom. In this treatment, all collisions are assumed to be inelastic and collision rates used are derived from experimental data. In order to determine the feasibility of sensing hydrogen and helium using acoustic methods a model based on molecular relaxation\cite{petculescu2004fine} is constructed. 

\par Molecules are free to move (translate) as well as vibrate or rotate. There are two distinguished contributions due to acoustic losses: 1) the classical attenuation arising from viscous friction and thermal conduction, and 2) the non-classical (or relaxational) attenuation due to the inability of the internal motion (generally vibration and rotation) of molecules to follow the acoustic fluctuations.

\par 
Generally, the period of a sound wave is significantly shorter than the time needed to establish thermal equilibrium with the surroundings, which leaves the wave to propagate adiabatically. As the compressional part of the sound wave interacts with a gas, a local increase of temperature occurs. This encourages molecules to occupy higher energy states by upholding a Boltzmann distribution. The excitation of the molecules to higher energy states can only occur during collisions; therefore, it takes a characteristic time for the gas to adjust to the temperature change. This results in a phase lag between the relaxing degrees of freedom and the ambient temperature. Each internal mode of the molecule will respond to changes in temperature at a different rate. Thus, each internal mode has its own characteristic relaxation time.
\par For hydrogen, vibrational excitation occurs at temperatures of 3400 K. This is beyond the range for gas-giant atmospheric conditions that any proposed entry probe would experience. Conversely, rotational excitation can occur at temperatures starting at 88 K. Moreover, due to the inverse relationship between rotational energy spacing and mass, hydrogen is a prime candidate for rotational spectroscopy.  Therefore, an investigation of the rotational relaxation of hydrogen is pursued. Additionally, the presence of helium affects the relaxation times of hydrogen and is also included\cite{tejeda2008low}. Collisions between hydrogen and helium are pure kinetic exchanges, as the helium is a non-relaxing gas.

\subsection{Relaxation times}
\par 
Due to the finite nature of energy transfer, an ``excited" molecule takes an amount of time to relax back to the ambient temperature. This time is classified as the relaxation time given by\cite{herzfeld1959absorption}:
\begin{equation}
    \frac{dT_{r}}{dt} = -\frac{(T_{r}-T)}{\tau}
\end{equation}
where $T_{r}$ is the temperature at which the relevant relaxation process occurs, T is the translational/ambient temperature, and $\tau$ is called the relaxation time. The rotational excitation of a two molecule system, $M_{a}, M_{b}$, has a collisional process described with:
\begin{equation}
    M_{a}(i) + M_{b}(j) 
\stackrel{k(T)_{ij\rightarrow lm}}{\longrightarrow} M_{a}(\ell)+M_{b}(m)
\end{equation}
where the molecules exchange pre-collisional quantum states, (i,j), with those after the collision, ($\ell$,m).The relevant collisions are oH2:oH2, oH2:pH2, and pH2:pH2. Despite the absence of internal degrees of freedom in helium, it affects the relaxation times of hydrogen through He-H$_2$ collisions, as described in \cite{tejeda2008low}.
\par Using the treatment \cite{montero2014rotational}, the relaxation time is calculated as:
\begin{equation}
\begin{split}
    \tau^{-1} = \frac{nk_{b}}{c_{rot}}\times \Biggl[ & \alpha_{a}\alpha_{a} \sum_{i}\sum_{j}\sum_{\ell \leq m}\sum_{m} Q_{ijlm}(\mathcal{E}_{i}+\mathcal{E}_{j}-\mathcal{E}_{\ell}-\mathcal{E}_{m})^{2} P_{\ell}P_{m}k_{\ell m \rightarrow ij} \\
    & +2\alpha_{a}\alpha_{b}\sum_{s} \sum_{i}\sum_{u}\sum_{l}(\mathcal{E}_{s}+\mathcal{E}_{i}-\mathcal{E}_{u}-\mathcal{E}_{\ell})^{2} P_{u}P_{\ell}k_{u\ell\rightarrow si} \\
    & +\alpha_{b}\alpha_{b}\sum_{r}\sum_{s}\sum_{t\leq u}\sum_{u}Q_{rstu}(\mathcal{E}_{r}+\mathcal{E}_{s}-\mathcal{E}_{t}-\mathcal{E}_{u})^{2}P_{t}P_{u}k_{tu\rightarrow rs}\Biggr]
\end{split}
\label{eq:mont}
\end{equation}
where $n$ is the number density obtained from the ideal gas equation of state as $n = p/k_{b} T$. The detailed-balance relation between up and down state-to-state rates is:
\begin{equation}
    k_{\ell m\rightarrow ij} = k_{ij\rightarrow \ell m}\frac{(2i+1)(2j+1)}{(2\ell+1)(2m+1)}e^{(E_{\ell}+E_{m}-E_{i}-E{j})/k_{b}T_{r}}
\end{equation}
and $E_{i}$, is the rotational energy given from $E_{i} = i(i+1)\hbar^{2}/2I$, with I being the moment of inertia, $\hbar$ is the reduced plank's constant and $\mathcal{E}_{i} = E_{i}/(k_{b}T)$. The quantity $Q_{i\ell j m}$ is given by:
\begin{equation}
    Q_{ijlm} = [1+\delta_{ij}(1-\delta_{\ell i})(1-\delta_{mi})]\times[1-\delta{\ell i}(1-\delta_{ij}]\times[1-\delta_{mi}(1-\delta_{ij}]
\end{equation}
is used to avoid double counting indistinguishable particles. The populations are obtained from $P_{i}$,  and refer to the population of a particular rotational level i. Furthermore, $\alpha_{a}, \alpha_{b}$ are the mole fractions of the respective molecules a,b. Finally, the rotational specific heat capacity is given by $c_{rot}$ and $k_{b}$ is Boltzmann's constant. 
Thus for the case where collisonal rates are available, the calculation of the relaxation time of $H_{2}$ can be calculated from \ref{eq:mont}. 


\subsection{Acoustics}
Perturbations from the passing of an acoustic wave cause the gas to re-establish equilibrium with respect to the ambient background. Absorption from the wave is bounded by the set relaxation time $\tau$. 
\par In a multi-component gas mixture, the combination of thermo-viscous coupling and relaxation processes predict a frequency dependent effective wave number\cite{pierce2019acoustics}:
\begin{equation}
    \tilde{k}(f) = 2\pi f \sqrt{\frac{\rho_{0}}{p_{0}}\frac{\tilde{C}_{V}(f)}{\tilde{C}_{P}(f)}} 
\end{equation}
Derived from this equation is the sound speed and attenuation.
\begin{equation}
    c = \frac{2\pi f}{\mathrm{Re}[\tilde{k}]} \qquad \alpha = \mathrm{Im}[\tilde{k}] + \alpha_{\mathrm{class}}
\end{equation}

where $c$,$\alpha$ are the sound speed and attenuation respectively. The classical attenuation is given by $\alpha_{\mathrm{class}}$ which is due to thermal, viscous, and diffusional transport. The density and pressure are given by $\rho_{0}$,$p_{0}$ and $\tilde{C}_{P}$,$\tilde{C}_{V}$ are the isobaric and isochoric specific heats. 
The classical attenuation coefficient is given from\cite{beyer}:
\begin{equation}
    \alpha_{\mathrm{class}} = \frac{2\mu f^{2}}{3\rho_{0}c^{3}}+\frac{\gamma - 1}{C_{P}}\frac{f^{2}\kappa}{2\rho_{0}c^{3}}
\end{equation}
The mixture averages of Wilke\cite{poling2001properties} are used to generate the viscosity $\mu$ and thermal conductivity $\kappa$, and $\gamma$ is the specific heat ratio. 
\\
The effective wave number can be show to be:
\begin{equation}
    k^{2} = \frac{\omega}{RT}\frac{C_{V} + (\mathrm{x}H_{2}(\Gamma - 1))c_{rot,m}}{C_{P} + (\mathrm{x}H_{2}(\Gamma - 1))c_{rot,m}}
\end{equation}
where the lapse parameter $\Gamma$ is given as:
\begin{equation}
    \Gamma = \frac{1}{1-i\omega \tau}
\end{equation}
the angular frequency is $\omega = 2\pi f$ and $c_{rot,m} = c_{rot}/m$ represents the rotational specific heat divided by the mass of the relaxing molecule. 

From equations (9) and (8) we can arrive at equation (7), thus generating acoustic quantities relevant to sensing. 
\section{Results}
\subsection{Model Validation}
As a preliminary check, the model is compared against the seminal results of \cite{Raff1968} Here ultrasonic experiments are compared at unit conditions, i.e. room temperature and 1 atmosphere. Included in the model are conversions for the 1 bar case presented. The two relevant quantities reported are the sound speed and $\alpha \lambda$. It is important to note that contained within \cite{Raff1968}  are purely the relaxational contributions to $\alpha \lambda$.
\begin{figure}[htbp]   
\centering
\includegraphics[scale=0.55]{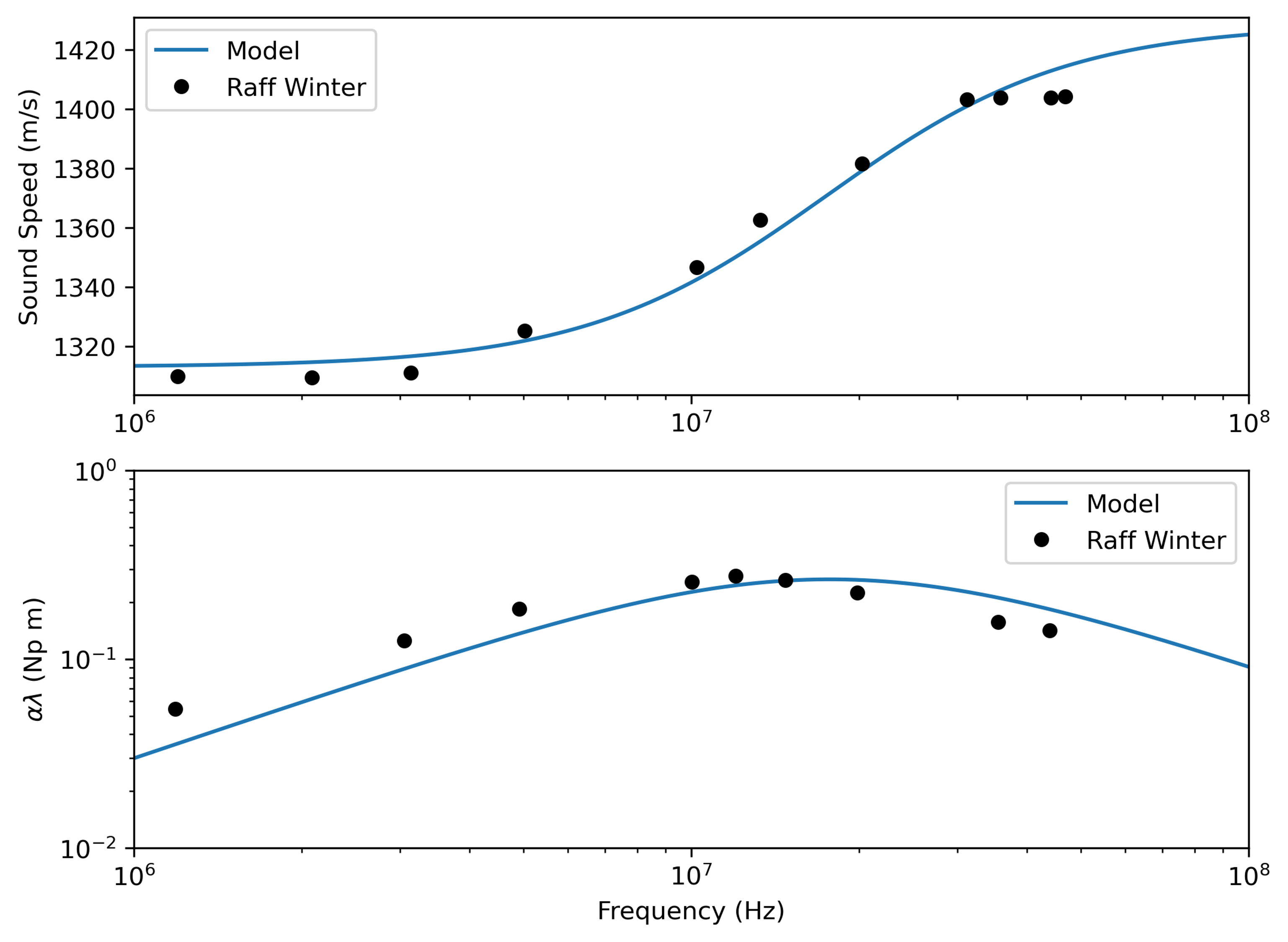}
\caption{Digitized data from \cite{Raff1968} compared to model at inputs of 1 bar and 295K.}
\label{fig:rw}
\end{figure}
\\
As seen from Figure \ref{fig:rw} the model and experiments are in relative agreement. Possible errors include the presence of humidity within the chamber of the experiment, as well as small potential errors in the measurement of pressure. These results serve to bolster the models confidence and therefore more scenarios, particularly those relevant to atmospheric sensing, are pursued. 

\subsection{Temperature profile}
As a preliminary step, the profiles of \cite{KOSKINEN2018161} were used to represent the characteristic temperature and pressures experienced by a prospective entry probe. 

\begin{minipage}{\textwidth}
\centering
\includegraphics[scale=0.5]{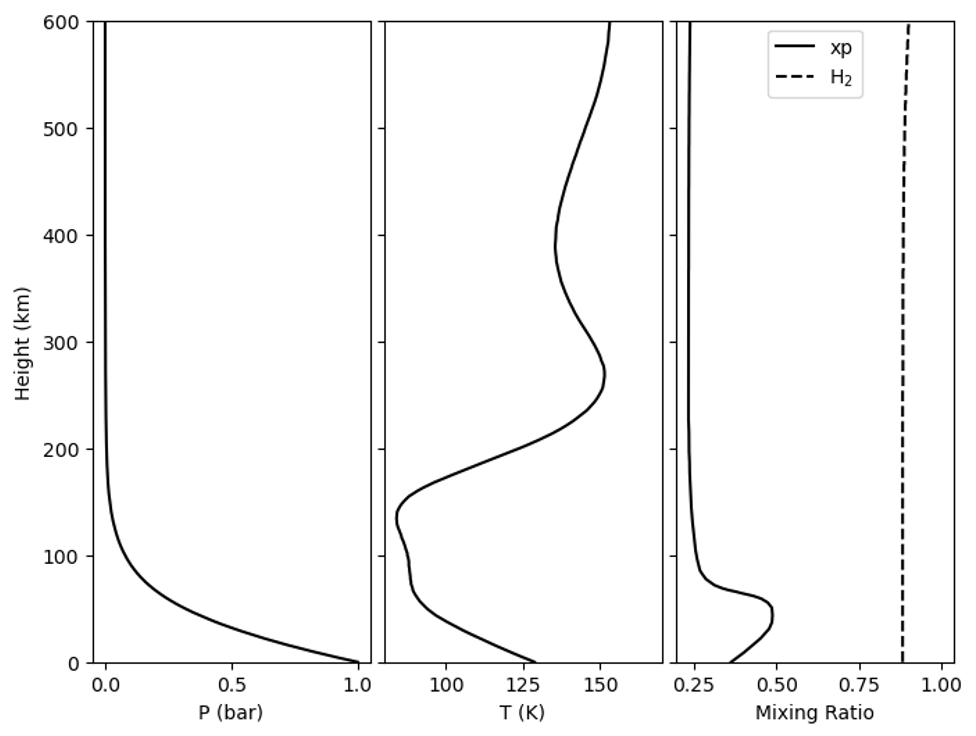}
\captionof{figure}{The temperature, pressure, and mixing ratio profiles extracted from \cite{KOSKINEN2018161}}
\label{fig:thermo_phys}
\end{minipage}

The pressure follows the scale height relation as:
\begin{equation}
    p = p_0 e^{-z/H}
\end{equation}
where $H = RT/Mg$ represents the scale height. For Saturn the scale height is approximately: $H \sim 59.5 $km.

At each altitude, characterized by a temperature-pressure pair, thermophysical parameters such as $C_V$, $C_P$, $\kappa$, $\rho$, and $\mu$ were extracted for both ortho/para-H$_2$ and He. Afterwards, varying combinations of chemical abundances were explored. This was done by generating different amounts of H$_2$ and its ortho/para fractions. From the inclusion of the thermophysicals combined with the configurations of abundances, the acoustic wavenumbers could be quantified and compared.  

\begin{minipage}{\textwidth}
    \centering
    \includegraphics[scale=0.5]{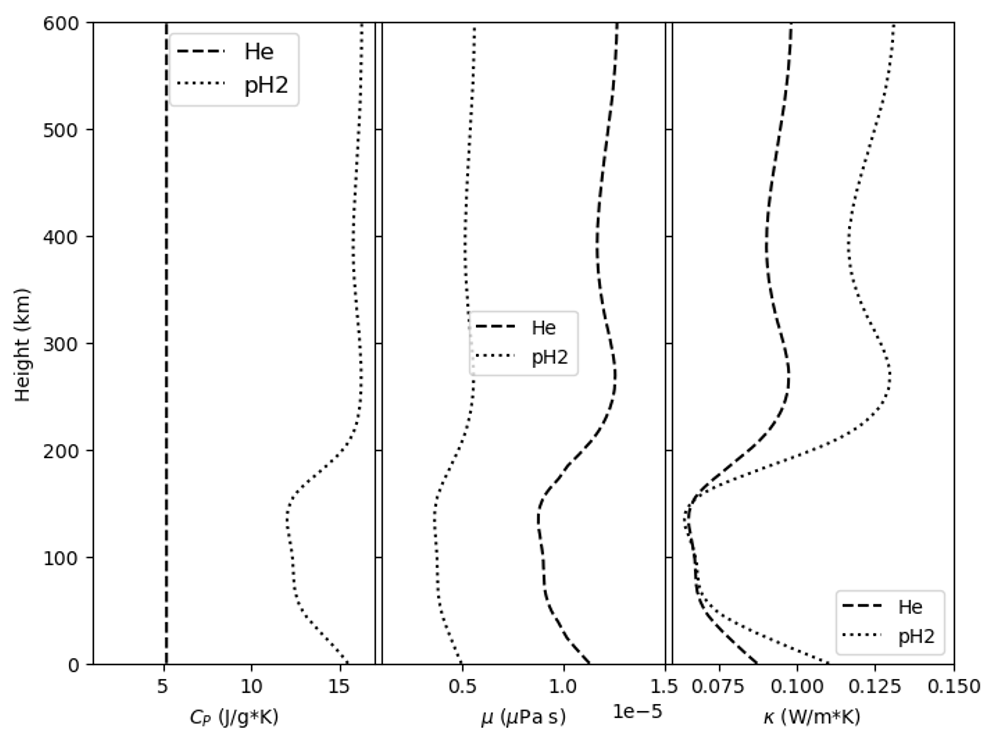}
    \captionof{figure}{Thermophysicals extracted at each t-p pair from Figure \ref{fig:t-p-profile} taken from NIST's Chemistry Webbook. Here only para-H$_2$ is provided.}
    \label{fig:enter-label}
\end{minipage}

\subsection{Sound Speed and Total Attenuation}
The rotational relaxation of the hydrogen molecules leads to a frequency dependence of the sound speed (intrinsic dispersion).
As the molecule becomes excited at a characteristic frequency that corresponds to $f_\text{char} = 1/2\pi\tau$, it gains energy from the wave causing a net increase in sound speed. For preliminary tests at room temperature (295 K), atmospheric pressure(1.0 bar), and a purely normal and homogeneous mixture of hydrogen($\mathrm{x}H_{2} = 1$) the ortho/para distribution is $\mathrm{x(p}H_{2}) = 0.25$ and $\mathrm{x(o}H_{2}) = 0.75$. is provided. The minimum speed before relaxation occurs is approximately $1312 ~\mathrm{m/s}$, whereas the maximum velocity obtained from the wave is $1423 ~\mathrm{m/s}$. Thus the dispersion accounts for a net increase of around $8.4\%$. 
\begin{figure}[htbp] 
\centering
\includegraphics[scale=0.75]{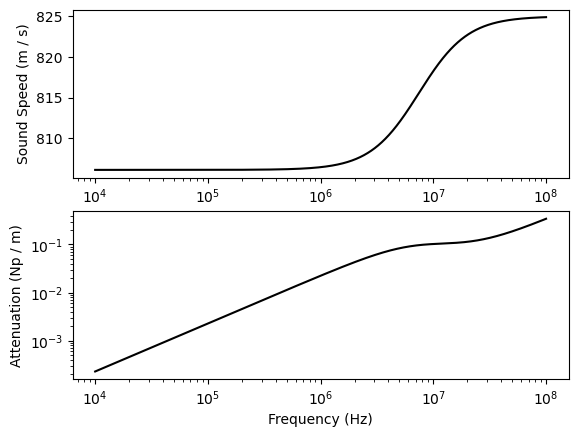}
\caption{Caption}
\label{fig:enter-label}
\end{figure}
\subsection{Classical and relaxational attenuation}
In order to emphasize the effects of the relaxation portion of the attenuation. The vertical axis is scaled by the sound speed divided by the frequency. The resulting curve is thus $\alpha \lambda$. The local maximum of $\alpha\lambda$ occurs at the characteristic frequency $f_\text{char} \approx 17.4$ MHz, related to the rotational relaxation time, $\tau = (2\pi f_\text{char})^{-1} \approx 9.15$ ns.
This depletion of the energy from the wave allows for frequencies after the characteristic to be frozen in and thus flattening the attenuation curve. That is, rotational modes frozen in by the wave do not deplete additional energy from the wave and instead re-energize the mode frozen in as it relaxes. In Figure \ref{fig:alpha-lambda-total} the various contributions to the total attenuation are provided. Here, at lower frequencies the attenuation is dominated by the presence of relaxation phenomena. Due to the $f^2$ dependence on the classical attenuation, after the characteristic frequency, as the relaxation process continues to decline, the classical attenuation dominates at those high frequencies.
\begin{figure}[htbp] 
\centering
\includegraphics[scale=0.65]{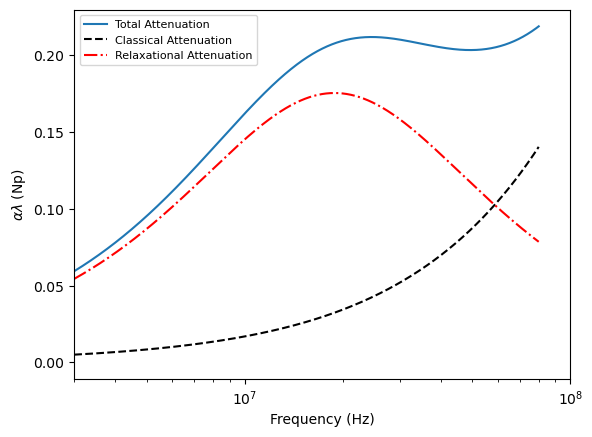}
\caption{The relative contributions to the total attenuation.}
\label{fig:enter-label}
\end{figure}
\subsection{Characteristic Frequencies}
In the table we present various scenarios, and the effect on the relaxation frequency of the mixture. The importance of the para ratio is evident in the blue shifting of the relaxation frequency. There is a similar effect in the presence of helium. The first four rows correspond to values relevant for room temperature and pressure testing. The last three rows are those of interest to a descent probe with the maximum frequencies around 2.66 MHz.  

\begin{minipage}{\textwidth}
\centering
\begin{tabular}{|c|c|c|c|c|}
\hline
$\chi_p$ & $\chi_{\mathrm{He}}$ & T(K) & P(bar) & f(MHz) \\ 
\hline
    0.25 & 0& 295& 1& 17.5 \\ 
    0.50 & 0& 295& 1& 21.4 \\ 
    0.25 & 0.1 & 295 & 1& 21.9 \\
    0.50 & 0.1 & 295 & 1& 24.5 \\
    0.33 & 0.12 & 88 & 0.15& 2.66 \\
    0.29 & 0.12 & 88 & 0.05 & 1 \\
     0.27 & 0.12 & 100 & 0.01 & 0.18 \\

\hline
\end{tabular}
\end{minipage}
\subsection{Sound speed and attenuation profile}
The temperature profile of the gas giants are given from UV occultation data of \cite{KOSKINEN2018161}. Whereas the former radio occultation data penetrated deeper in the atmosphere\cite{Lindal1985}, the UV occultation provides a higher resolution. Due to its $RT/M$ dependence, the sound speed profile closely mimics the temperature profile and displays the typical velocity a sound wave will encounter at various heights above 1 bar at a particular frequency. 

\begin{figure}[htbp!]
    \centering
    \includegraphics[scale=0.65]{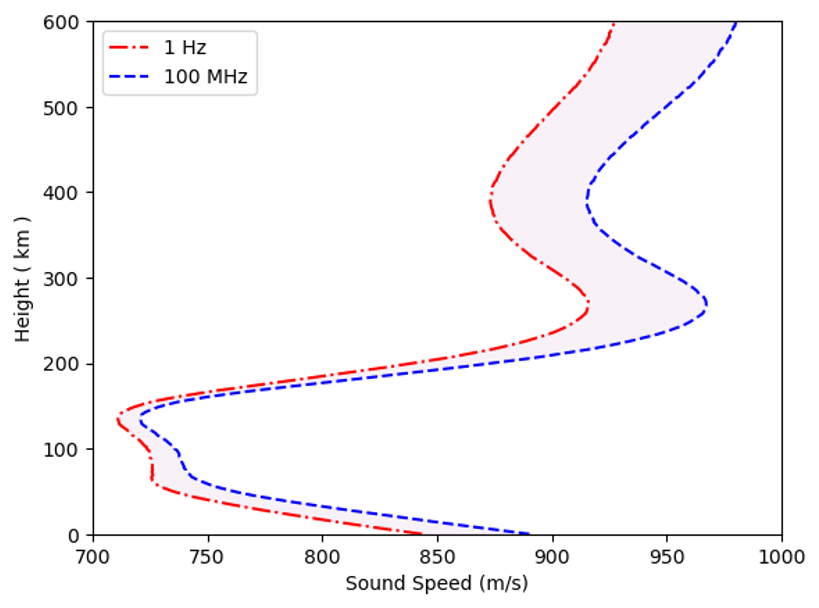}
    \caption{The sound speed profile of a selected occulation of \cite{KOSKINEN2018161}}
    \label{fig:ss-profile}
\end{figure}
The attenuation profile represents the typical losses that a sound wave experiences through its descent in the atmosphere. Here, it is seen that clearly for higher frequency waves, the attenuation is greater, with its losses due to rotational relaxation being frozen in. This is also illustrative of the $1/f^2$ dependence of the classical attenuation. Within the 100 MHz regime is in the deep activation of both ortho and para hydrogen, thus measurements here are sensitive to those quantities. As the probe descends through the atmosphere, due to the dominance of the high temperatures, the attenuation is highest.

\begin{figure}[htbp!]
    \centering
    \includegraphics[scale=0.65]{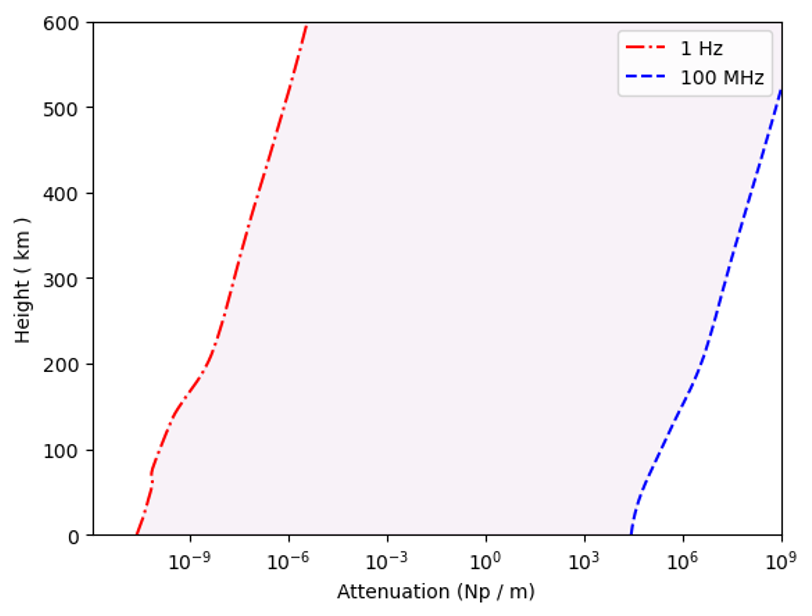}
    \caption{Attenuation profile of a selected occultation of \cite{KOSKINEN2018161}}
    \label{fig:att-profile}
\end{figure}
\section{Conclusions}
Contained within this work has been an initial attempt to model the expected acoustic signatures in a gas giant atmosphere, the preliminary body being Saturn. Using a combination of the frequency-dependent complex wavenumber from linear acoustics and the rotational relaxation time, obtained from molecular jet experiments, we predict the characteristic frequencies that are relevant for any instrument aiming to probe the molecular abundances using acoustic methods. The primary region of interest for an entry probe is on the order of 50 mbar to 150 mbar. Here the frequencies range from \textcolor{red}{1} MHz to \textcolor{red}{2.66} MHz, which are feasible for broadband ultrasonic transducers. The depletion of helium, in particular, has prominent implications for Saturn's upper atmosphere and may prove to yield insight into sequestration processes deep within the interior of the planet. The ortho/para ratio emerges as a consequence of its different rotational transitions, which are more apparent at higher frequencies, but still may be probed in the range of interest. 

Further examination of alternate profiles could generate a more global model, thus increasing the chances of detection at more promising entry latitudes. Here we take an equatorial profile and examine those atmospheric conditions. Using the planet's tropics or polar occultations may yield insight into areas of the atmosphere that are more driven by vertical transport, rather than solar radiation. This would prompt a more sensitive situation for the ortho/para ratio, which has an influence on the atmospheric stability. 

Alternately, an extension of the model could be applied to other giant planets such as Jupiter, Uranus, or Neptune. Here, although the helium depletion is less, a markedly apparent impact in its acoustic signatures could be observed. Furthermore, any gas-giant exoplanet candidate with a proposed/estimated thermophysical profile would be able to garner treatment similar to that in the solar system. 

With all considerations, the use of acoustic measurements could help complement other elemental and molecular abundance investigations. Particularly, the estimated sound speed and attenuation profiles at selected T-P pairs could provide insight into longstanding questions of the helium abundance, which is critical for the planet's time of formation, and its ortho/para ratio, which can dictate the vertical transport of thermal energy. Future iterations of the constructed model could help generate a synthetic dataset suited for inference on measurements obtained from an atmospheric instrument. 

Continued comparison to experimental efforts of \cite{Rishabh_AIAA} serve to validate the model assumptions and practicality. Furthermore, as more measurements are made, transducer ranges can be better known. With advances in transducer construction and capabilities, higher pressures could be tested.

\newpage
\bibliography{main}
\bibliographystyle{plain}
\end{document}